\documentstyle[12pt, aaspp4]{article}

\lefthead{Coronal Holes at the 23rd solar minimum }
\righthead{Abramenko et al.}

\begin{document}

\title{Low-Latitude Coronal Holes at the Minimum of the 23rd Solar Cycle}

\author{Valentyna Abramenko and Vasyl Yurchyshyn}
\affil{Big Bear Solar Observatory, Big Bear City, CA 92314}

\author{Jon Linker and Zoran Mikic}
\affil{Predictive Science, Inc., San Diego, CA 92121}

\author{Janet Luhmann and Christina O. Lee}
\affil{Space Sciences Laboratory, University of California, Berkeley, CA 94720}

\begin{abstract}
Low and mid-latitude coronal holes (CHs) observed on the Sun during the current
solar activity minimum (from September 21, 2006, Carrington rotation (CR) 2048,
until June 26, 2009 (CR 2084)) were analyzed using {\it SOHO}/EIT and STEREO-A
SECCHI EUVI data. From both the observations and Potential Field Source Surface
(PFSS) modeling, we find that the area occupied by CHs inside a belt of
$\pm$40$^\circ$ around the solar equator is larger in the current 2007 solar
minimum relative to the similar phase of the previous 1996 solar minimum. The
enhanced CH area is related to a recurrent appearance of five persistent CHs,
which survived during 7-27 solar rotations. Three of the CHs are of positive
magnetic polarity and two are negative. The most long-lived CH was being formed
during 2 days and existed for 27 rotations. This CH was associated with fast
solar wind at 1 AU of approximately 620$\pm 40$ km s$^{-1}$.  The 3D MHD
modeling for this time period shows an open field structure above this CH. We
conclude that the global magnetic field of the Sun possessed a multi-pole
structure during this time period. Calculation of the harmonic power spectrum of
the solar magnetic field demonstrates a greater prevalence of multi-pole
components over the dipole component in the 2007 solar minimum compared to the
1996 solar minimum. The unusual large separation between the dipole and
multi-pole components is due to the very low magnitude of the dipole component,
which is three times lower than that in the previous 1996 solar minimum.
\end{abstract}


\section {\bf Introduction}

Coronal holes (CHs), in general, may be observed anywhere on the solar disk,
however their spatial distribution, frequency of occurrence and evolution are
governed by the solar cycle. Three types of CHs may be distinguished (Storini et
al., 2006). First, there are {\it polar} CHs, which are located at the solar
poles and have a lifetime comparable to that of the solar cycle.  Included in
this category are CH extensions, such the ``elephant trunk'' CH observed in
August 1996 (Zhao et al.\ 1999). Second type is {\it isolated} CHs, which are
mostly limited to the low- and middle latitudes and are not connected to the
polar CHs. Their lifetime is comparable to few Carrington rotations (CRs). The
third type consists of {\it transient} CHs, which are short lived (hours or
days) and their occurrence is often associated with powerful eruptions (flares)
and/or filament disappearances. 
Polar CHs usually disappear during the maximum of a solar cycle and reappear
again, already having magnetic flux of opposite polarity, thus manifesting the
polar field reversal. Isolated CHs behave quite differently. They tend to be
more often detected at activity maximum and disappear near activity minimum.
According to Hofer \& Storini (2002), during the 1986 solar minimum there were
only few isolated CHs and none of them was older than 5 solar rotations. Similar
findings were reported by Bilenko (2002) for the 1996 minimum. Potential Field
Source Surface (PFSS) modeling of open field regions (Luhmann et al.\ 2002) also
indicates a very low population of low- and mid-latitude CHs during the two
previous (1986 and 1996) solar minima. 

Qualitatively, the above reports are consistent with the comparison of the
structure of the solar corona at the 1996 solar minimum derived from 3D MHD
modeling (Miki{\'c} et al.\ 1996; 1999; Linker et al.\ 1996; 1997; 1999) and
actually observed during two solar eclipses in 1995 and 1997 (Figure
{\ref{fig1}). Apparently, the lack of isolated low-latitude CHs on the solar
surface resulted in the formation of a well-defined closed magnetic
configuration near the solar equator (streamer belt) and the well developed {\it
dipole} structure of the global magnetic field of the Sun.

\begin{figure}[!h] \centerline {
\epsfxsize=3.0truein \epsffile{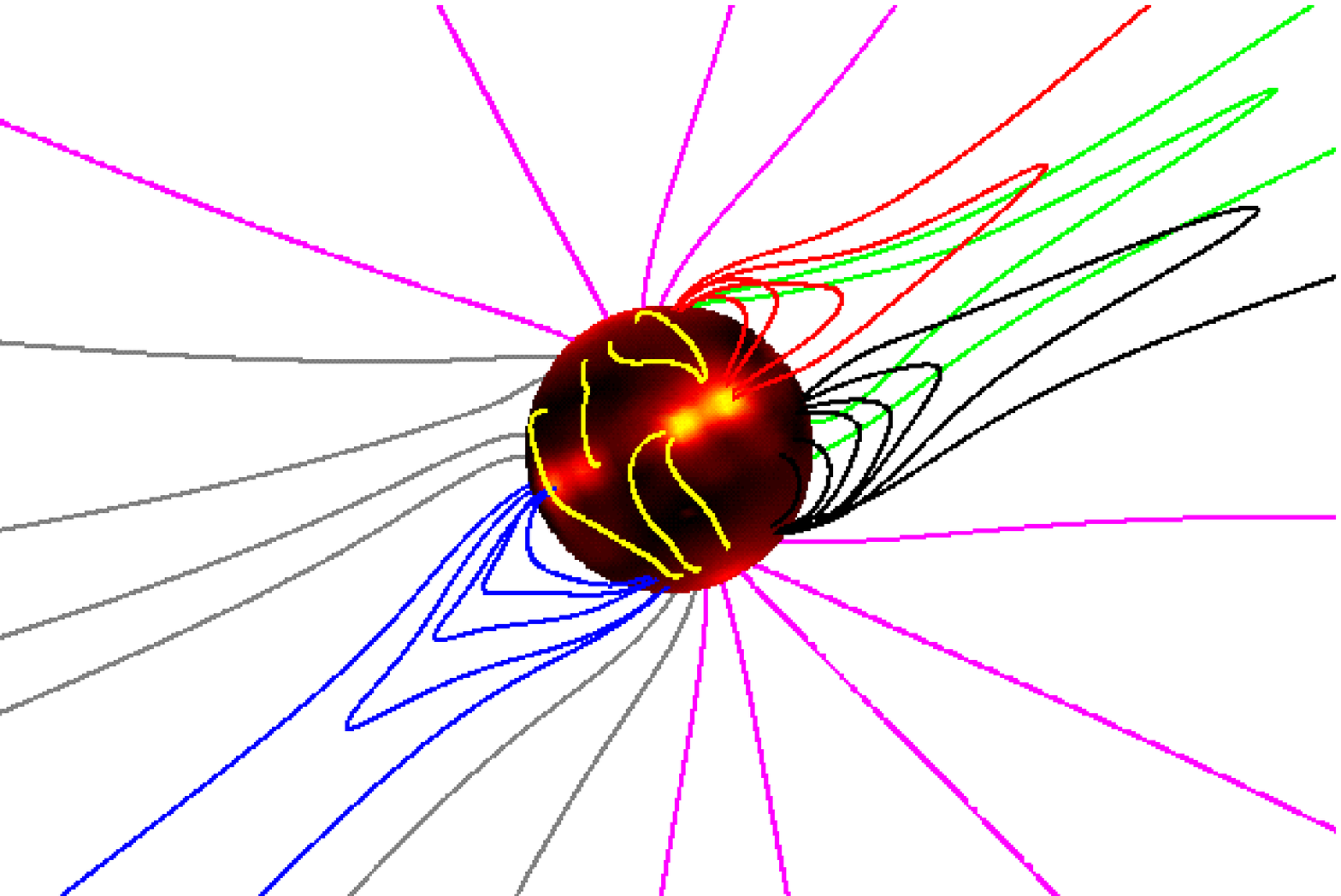}
\epsfxsize=3.0truein \epsffile{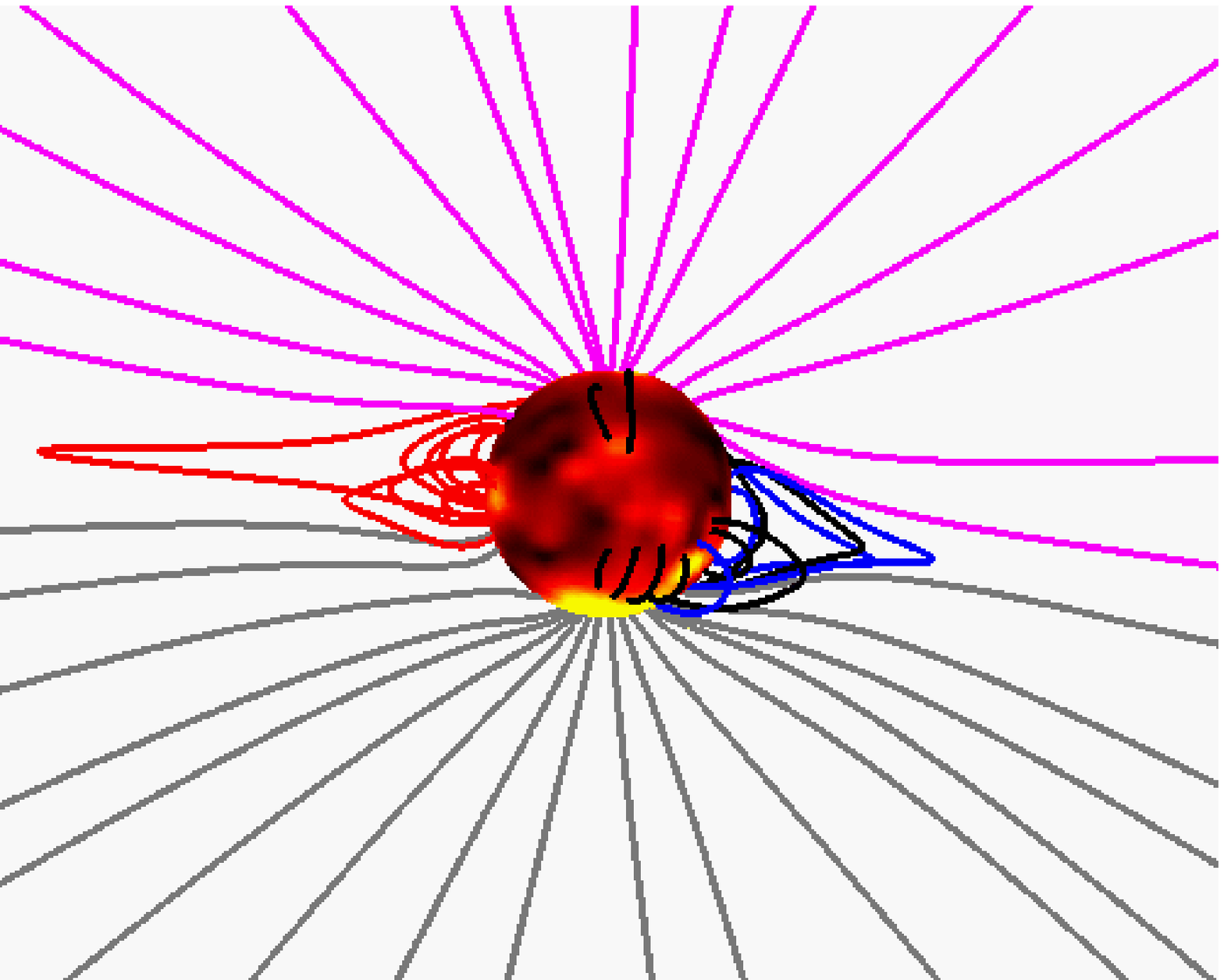}}
\caption{\sf A typical structure of the solar corona at a solar minimum.
{\it Left - } Magnetic field lines in the corona, shown in arbitrary colors, as
derived from a 3D numerical simulations for CR1900 prior to the October 24, 1995
solar eclipse. The photospheric magnetic field data used in this model are
Wilcox Solar Observatory synoptic maps. The image in the photosphere shows the
intensity of the magnetic field: active regions with strong magnetic fields
appear yellow. The view point corresponds to the approximate time of the eclipse
(Carrington longitude of 39 degree at the central meridian). Terrestrial north
is to the top. {\it Right - } The same as the left panel but for the March 9,
1997 solar eclipse (Carrington longitude of 257 degrees). Note that in both
models the dipole structure of the global magnetic field is very prominent and
there are no significant inclusions of open flux in the equatorial zone.}
\label{fig1}
\end{figure}

The above scenario seems not to be applicable for the current extended 2007
solar minimum, when many global parameters of the heliospheric magnetic field
differ significantly from what was measured in the past solar minima. Apart
from the unusually low measured sun spot number, observations also show that: i)
interplanetary magnetic field strength is lower when compared to a similar phase
of the previous 22nd solar cycle (Lee et al.\ 2009); ii) the meridional flows
inferred by means of surface feature tracking are slower (Hathaway 2008,
Schrijver \& Liu 2008); iii) photospheric polar magnetic flux is smaller (Smith
\& Balogh 2008; Lee et al.\ 2009); iv) solar wind mass flux dropped by about
20\% (McComas et al.\ 2008); v) the tilt angle of the solar dipole throughout
the current 2007 solar minimum is unusually high and vi) co-rotating interactive
regions (CIRs) are abundant and geomagnetic activity is remarkably persistent
and recurrent (Mason et al.\ 2009).

Recently, a comprehensive analysis of a part of the declining phase of the 23rd
solar cycle (from January 8 to November 4, 2007) using the PFSS and
Wang-Sheeley-Arge (WSA, Arge \& Pizzo 2000) models and comparison with STEREO
and OMNI solar wind data has been carried out by Luhmann et al.\ (2009) and Lee
et al.\ (2009). The model results clearly demonstrated an excess of low- and
mid-latitude CHs during the time period under study (Figure 8 in Luhmann et al.\
2009) and show a reduced magnitude of solar polar magnetic field. The presence
of low- and mid-latitude CHs may have affected the enhanced ecliptic solar wind
speed and explain the observed anomalous distribution of the solar wind speed
with a heavy tail at 500-700 km s$^{-1}$.

In this paper, we take a closer look at the low-latitude CHs during the period
between 2007 and 2009 by utilizing data from the {\it Solar and Heliospheric
Observatory}  (SOHO)  {\it Extreme ultraviolet Imaging Telescope} (EIT,
 Delaboudiniere et al.\ 1995) and {\it Solar Terrestrial Relations Observatory}
(STEREO) {\it Sun Earth Connection Coronal and Heliospheric Investigation}
(SECCHI) EUVI (Howard et al. 2008) instruments. We investigate their recurrent
properties by tracking CHs in time and we analyze the harmonic power spectrum of
the solar magnetic field (Section 2). In Section 3 we analyze one of the coronal
holes in more detail, including its magnetic field structure and solar wind
signatures. Concluding remarks are in Section 4.

\begin{figure}[!h] \centerline{ \epsfxsize=5.0truein \epsffile{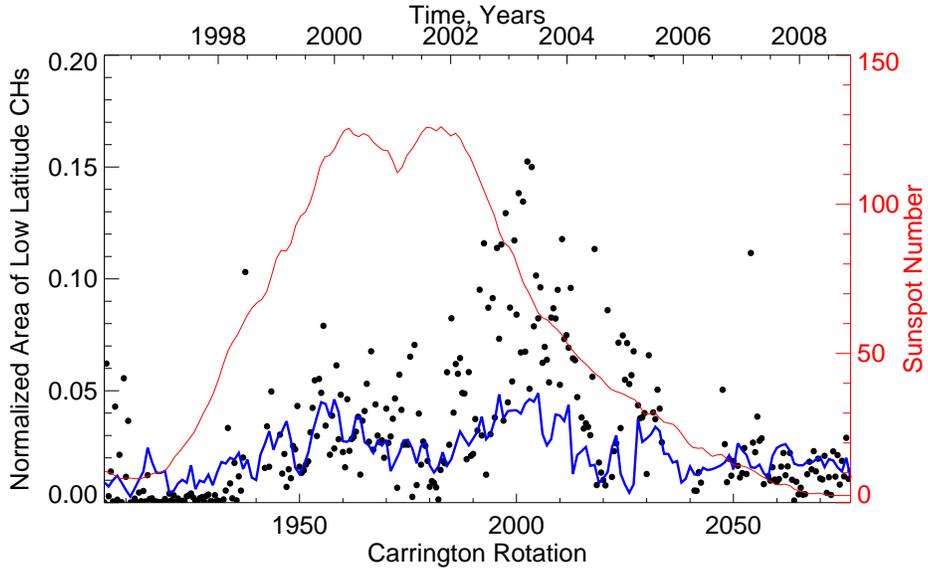}}
\caption{\sf {\it Dots -} The total area of the low-latitude (-40$^\circ$ to 
+40$^\circ$) CHs plotted versus the CR number (bottom horizontal axis, the top
horizontal axis shows years). The CH area was measured from SOHO/EIT Fe XV
284\AA\ images for each half of a CR and is plotted here normalized to the solar
disk area bounded between -40$^\circ$ and +40$^\circ$ of solar latitude. {\it
Blue -} the same area derived from the PFSS model with the source surface at 2.5
solar radii. {\it Red -} the sunspot number (right vertical axis). During the
current solar minimum (2006-2009) low-latitude CHs occur more frequently and
they occupy an area larger when compared to the previous solar minimum
(1995-1997).}
\label{fig2}  \end{figure}

\section {\bf High Population of Low-Latitude CHs during the Minimum of
Solar Cycle 23}

To study the occurrence frequency of low-latitude CHs on the solar disk
we calculated the total area of CHs (Figure \ref{fig2}) observed between
February 1996 (CR 1905) and December 2008 (CR 2077). The CH area was determined
from SOHO/EIT Fe XV 284\AA\ synoptic maps
($http://sun.stanford.edu/synop/EIT/index.html$) for all CHs located between the
latitudes of -40$^\circ$ and +40$^\circ$. This area was calculated for each half
of one Carrington rotation and then normalized to the total solar surface area
limited by -40$^\circ$ and +40$^\circ$ latitudes. (For details on the CH
outlining technique, see Abramenko et al.\ 2009). Correspondingly, the
normalized CH area was calculated from the PFSS model data (blue curve in Figure
\ref{fig2}) in the manner as described in Lee et al. (2009). As it follows from
the graph, both observed and model data show that, in contrast with the previous
1996 solar minimum, low-latitude CHs appear more frequently during the 2007
solar minimum and they occupy a larger portion of the solar surface.

\begin{figure}[!h] \centerline {\epsfxsize=6.5truein \epsffile{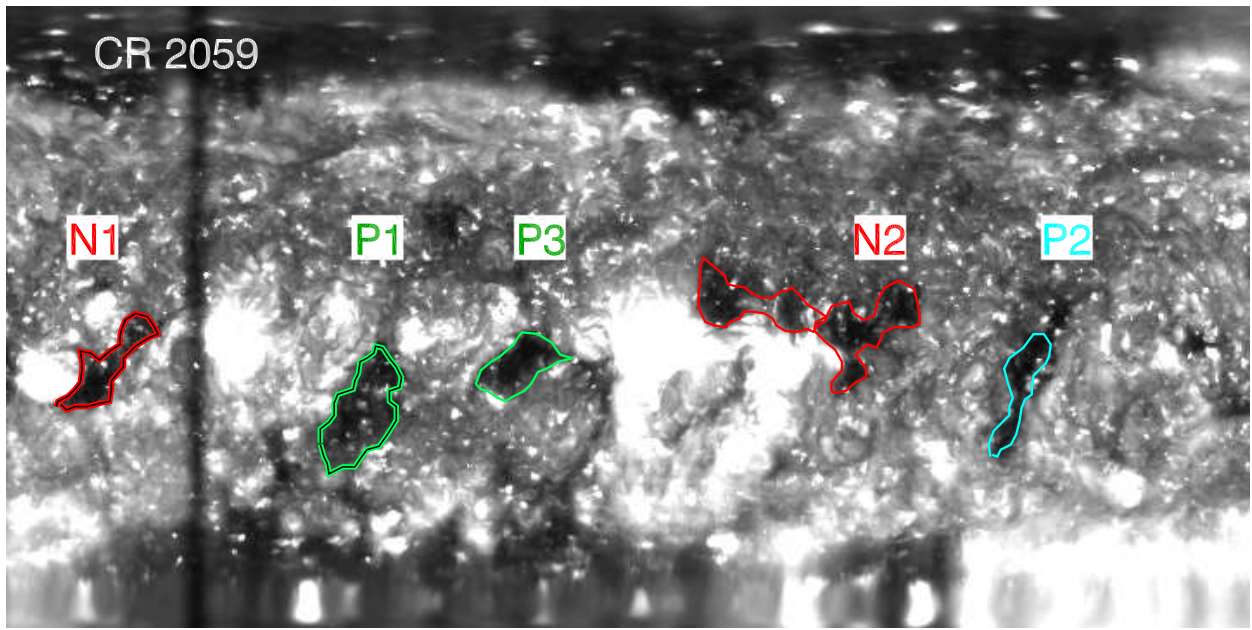}}
\caption{\sf STEREO-A SECCHI EUVI 195\AA\ synoptic map for CR 2059 (from July 18
to August 14, 2007) showing five low-latitude CHs. The red lines outline two
negative polarity CHs (N1 and N2), while the green and blue lines encircle three
positive polarity CHs (P1, P2 and P3).}
\label{fig3} \end{figure}

\begin{figure}[!h] \centerline{ \framebox{\epsfxsize=6.5truein \epsfbox{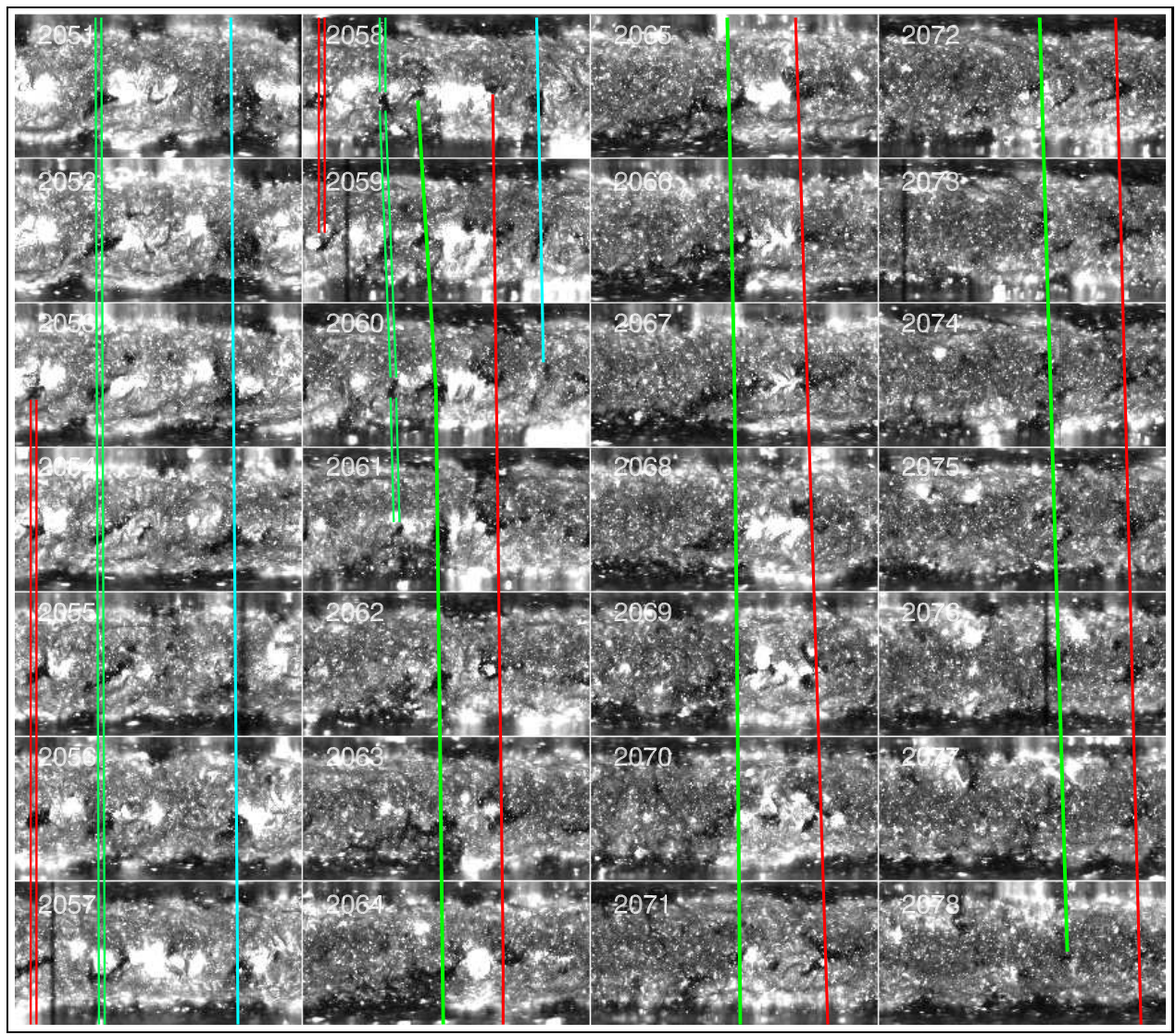}}}
\caption{\sf STEREO-A SECCHI EUVI 195\AA\ synoptic maps for 28 CRs ranging from
CR 2051 (December 12, 2006) to CR 2078 (January 12, 2009). Each color lines
traces the same CH at subsequent rotations and it crosses a map from top to
bottom in case when the corresponding CH existed during previous and subsequent
CRs. The lines have a gap to make small CHs visible. The red lines trace two
negative polarity CHs: N1 (double red line), appeared at CR 2053 and existed
during 7 rotations; N2 (thick red line), appeared at CR 2058 and was observed
during 27 rotations. The double green, blue and thick green lines trace positive
polarity CHs, P1, P2 and P3, respectively. The enlarged version of the synoptic
map for CR2059 is shown in Figure \ref{fig3}.} \label{fig4} \end{figure}

Are these low-latitude CHs short-lived transient entities (as they usually are
during a solar minimum), or is their appearance caused by structures
that survived many solar rotations? We address this question by tracking
the CHs in time.

In Figure \ref{fig3} we show a STEREO-A SECCHI EUVI 195\AA~ synoptic map
(background) where we outline the boundaries of five isolated low-latitude CHs.
Two of them were associated with negative polarity magnetic field (N1, N2) and
three were of positive polarity (P1, P2 and P3). Color lines in Figure
\ref{fig4} trace the CHs positions between consecutive Carrington rotations
(note that the color coding in Figures \ref{fig3} and \ref{fig4} is the same).
The positive polarity CHs were long-lived. P1 (double green line) and P2 (blue
line) appeared during CRs 2049 and 2048, respectively, and both lasted for about
13 solar rotations. Meanwhile, P3 (thick green line), which appeared during CR
2058, lasted for 21 solar rotations.

The negative polarity coronal hole N1 (double red line) appeared during CR2053
and was observed for at least 7 rotations (see also Luhmann et al.\
2009), while N2 (thick red line) was first detected during CR2058 (June 28,
2007) and survived until May 2009 for almost 27 solar rotations. Coronal holes
P3 and N2 observed during CR 2063 have been modeled by Luhmann et al. (2009).

It appears that the observed excess of low-latitude CHs during the current 2007
solar minimum is, to a large extent, caused by the repeated and persistent
appearance of these five CHs on the visible solar hemisphere. Quite possibly,
their existence may be a manifestation of strong multi-pole moments in the
harmonic spectrum of the magnetic field of the Sun, which can be written as
$S_n=\sum_{m=0}^n [(g_n^m)^2 + (h_n^m)^2]$ (Altschuler et al., 1977). Here,
$g_n^m$ and $h_n^m$ are harmonic coefficients determined from measurements of
the line-of-sight component of the photospheric magnetic field. We used the
harmonic coefficients calculated at the Wilcox Solar Observatory
($http://wso.stanford.edu/forms/prgs.html$) for CRs 2045 -- 2085 to examine the
dipole and multi-pole components during the 1996 and 2007 minima (Figure
\ref{fig5}). From these plots, the difference between the dipole and multi-pole
components is larger during the 2007 solar minimum as compared to the similar
phase of the 1996 solar minimum and it may be chiefly attributed to weakening of
the dipole component during the 2007 solar minimum. In particular, when
comparing data for CRs 1900 -- 1909 and CRs 2048-2057 (similar solar cycle
phase), one can see that the dipole component decreased nearly threefold,
whereas the sum of all multi-poles decreased only by about 30\%. It is also
interesting that there is a slight decline in both dipole and multi-pole
components during the 2007 solar minimum, which was not the case for the
preceding 1996 solar minimum. 

\begin{figure}[!h] \centerline {
\epsfxsize=3.5truein \epsffile{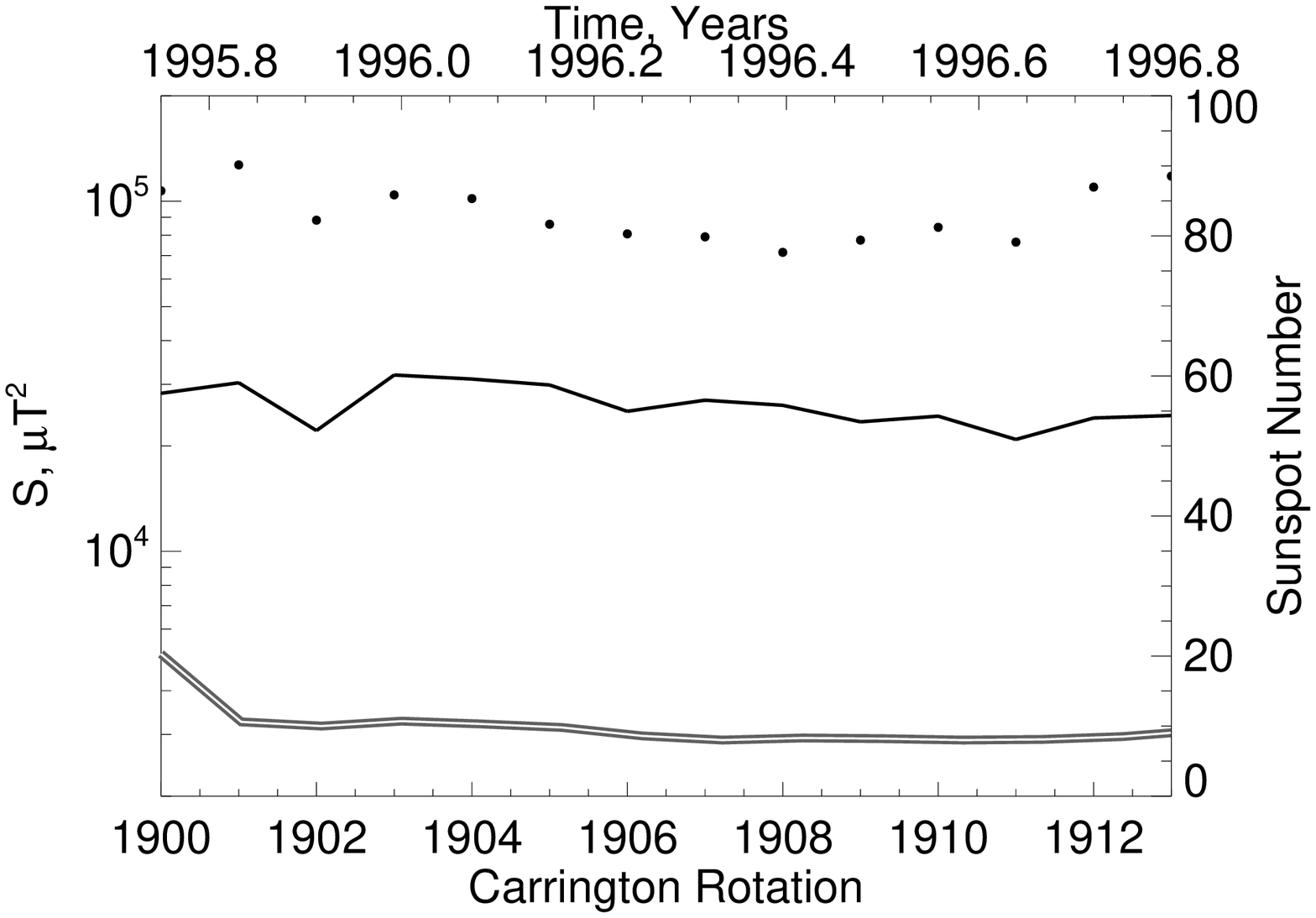}
\epsfxsize=3.5truein \epsffile{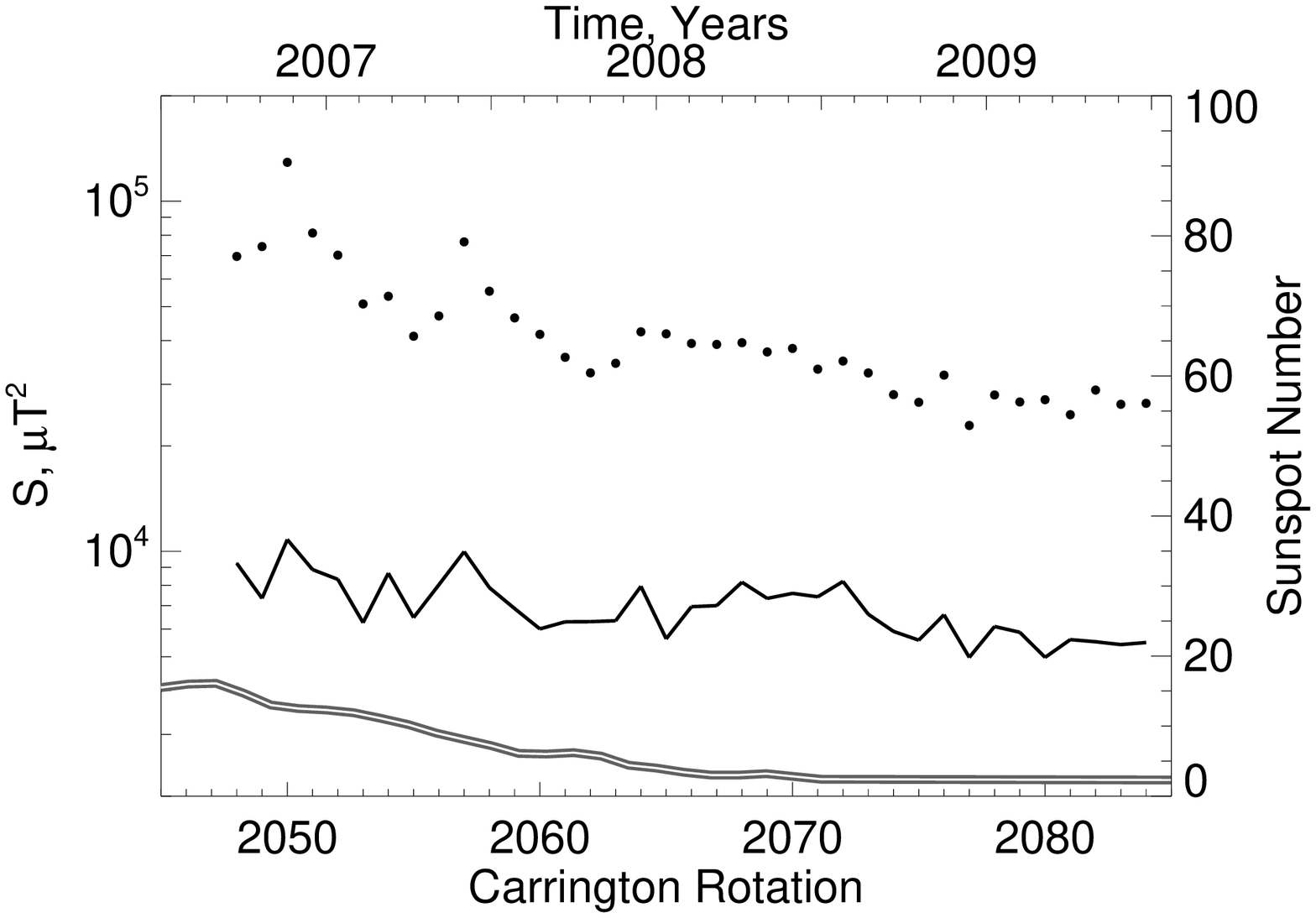}}
\caption{\sf Time profiles of various components, $S$ (in squared micro -Tesla),
of the multi-pole power
spectrum of the solar photospheric magnetic field during CRs 1900 to 1913 ({\it
left}) and CRs 2048 to 2085 ({\it right}). The solid line shows the dipole
component ($n=1$) and the dotted line are the sum of all multi-poles from $n=2$
to $n=7$. The sunspot numbers are plotted with double gray line (right axis).}
\label{fig5} \end{figure}

\section {\bf Genesis of the Equatorial Corona Hole N2}

We analyzed the most persistent coronal hole, N2, in more details. In Figure
\ref{fig6} we illustrate appearance of N2 by showing three EIT 195\AA\ images
taken on June 04, 28 and 30 of 2007. The over-plotted boxes encircle the same
area on the Sun. Comparison of the June 04 (top) and June 30 (bottom) images
(they are one CR apart) shows that small active region (AR) NOAA 10958,
completely decayed over one solar rotation and that a vast area immediately
north-east of the AR became occupied by a CH (indicated by the box in the bottom
panel). The June 28 EIT 195\AA\ image (middle) shows
that only two days earlier the CH was not present or, at least, it
was in a very early stage of development.

As we mentioned earlier, this CH was observed (having slightly different shapes)
for 27 solar rotations. The time difference between consecutive culminations of
the center of gravity of the CH was on average 26.9 $\pm$ 0.7 days. The latitude
of the center of gravity varied between -10$^\circ$ and 8$^\circ$, which
indicates on its near equatorial location. (Note that the coordinates of the
center of gravity were determined by averaging the corresponding coordinates of
the points located along the CH boundary; a technique to outline the CH boundary
is described in Abramenko et al.\ 2009). 

To probe the structure of the negative polarity field inside N2, we utilized
{\it Advanced Composition Explorer}
(ACE) {\it Solar Wind Electron, Proton, and Alpha Monitor} (SWEPAM) measurements
of the solar wind speed, $V_x$, in geocentric solar ecliptic (GSE) coordinate
system. We determined the speed of the fast stream, associated with the CH, by
accepting that the time interval between the stream arrival time at 1 AU and the
injection time (the CH culmination moment) multiplied by a given $V_x(t)$,
measured by ACE instrument, produces a distance of 1 AU (for more details see
Abramenko et al.\ 2009). This was performed for 17 solar rotations (CRs 2059 -
2075) during which N2 has been observed. Since no other solar activity was
detected during the period under study, our approach allowed us every time to
reliably detect a well-pronounced signature of the CH in the solar wind speed
time profile. The magnitude of the CH-associated solar wind speed varied between
560 and 670~km~s$^{-1}$ with the mean value of 620 $\pm$ 40~km~s$^{-1}$. 

We next examine whether there is evidence for the N2 CH in coronal models. A 3D
MHD model of the solar corona was computed prior to the August 1, 2008 solar
eclipse (see $http://www.predsci.com$). A synoptic magnetic field map for CRs
2071-2072 (June 25 -- July 21, 2008) was constructed from MDI magnetograms and
used as the boundary condition.  The modeling approach is described by Lionello
et al.\ (2009), while Rusin et al.\ (2009) describes a comparison of the model
with observations around the time of the eclipse.

Figure \ref{fig7} shows CH boundaries, magnetic field lines and emission at
195\AA\ as derived from the MHD model and the EUVI instrument aboard the
STEREO-B spacecraft on July 9, 2008. The open field regions in the model (black
areas and red field lines in Figure \ref{fig7}({\it a})) agree quite well with
the STEREO observations of N2 (Figure \ref{fig7}({\it c})). The simulated
emission (Figure \ref{fig7}({\it b})) also agrees reasonably well with the
observations, although the CH region is too dark and extensive in the model.
This is because the coronal heating mechanism, assumed for the model, produced
too little emission overall (see Lionello et al. (2009) for a discussing of the
technique for computing simulated emission from coronal models). Nevertheless,
it is clear that N2 is revealed to be a CH in the MHD model as well as the
observations. The N2 was detected again prior the August 2008 eclipse in
STEREO-A observations on July 14, and afterward in SOHO EIT observations on
August 7. It survived for nearly another year after that. During May, 2009 the
N2 began to shrink and it finally vanished one rotation later in June, almost
two years after its first appearance on the solar surface.

\begin{figure}[!h] \centerline {\epsfxsize=5.0truein \epsffile{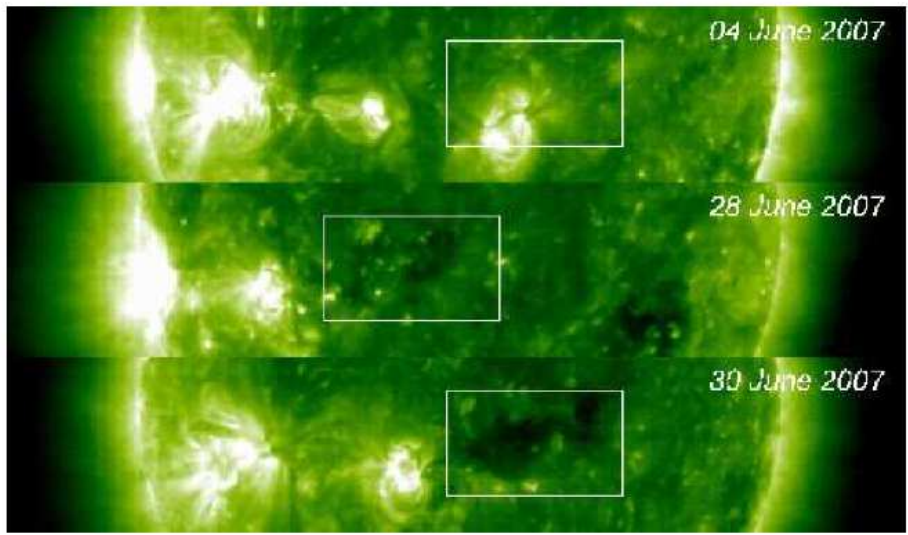}}
\caption{\sf Appearance of the coronal hole of negative polarity, N2, as
captured by EIT Fe XII 195\AA\ instrument. Boxes enclose the same area on the
Sun during two consecutive rotations on June 4 and 30 of 2007 (CRs 2057 and
2058). A small active region (NOAA 10958, inside the box in the top frame)
completely disintegrated and N2 formed on its place. Middle panel shows data on
June 28, which indicates that the CH fully developed in less than 2 days and
survived for almost 27 solar rotations.}
\label{fig6} 
\end{figure}

\begin{figure}[!h] \centerline {\epsfxsize=6.5truein \epsffile{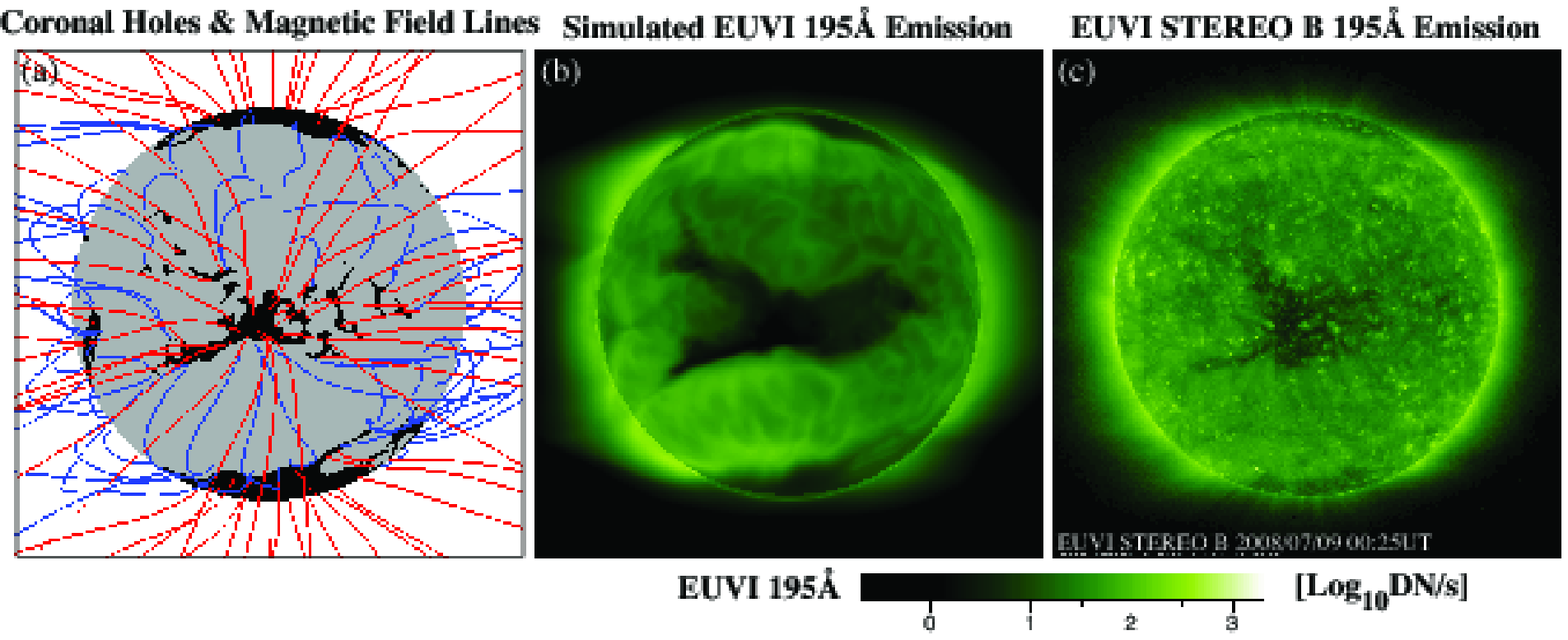}}
\caption{\sf {\it (Left -)} Coronal hole boundaries and magnetic field lines
from an MHD model of the solar corona prior to the August 1, 2008 solar eclipse.
Black regions shown on the surface are open field regions computed from the
model; gray areas are closed field regions. Closed magnetic field lines are
colored blue and open field lines are red. {\it Middle -} Simulated STEREO EUVI
195\AA\ emission from the MHD model and observed EUVI 195\AA\ emission from the
STEREO B spacecraft on July 9, 2008 ({\it right}).  Comparison of the three
images shows that the N2 coronal hole is clearly captured as an open field
region in the MHD model during this time period.} \label{fig7} \end{figure}

\section {\bf Concluding Remarks, Discussion and Open Questions}

We have combined observed data with magnetic field modeling to show that the
global magnetic field of the Sun during this 2007 extended solar minimum was not
a simple dipole. The area occupied by CHs inside a belt of $\pm$40$^\circ$
around the solar equator was found to be considerably larger during the 2007
minimum than that during the similar phase of the previous 1996 solar minimum.
Our results are confirmed by the CH-modeling results recently reported by
Luhmann et al.\ (2009).

Thus, the global magnetic field of the Sun had a multi-pole configuration, at
least in the time period between September 2006 and May 2009. The harmonic power
spectrum of the solar magnetic field shows a greater prevalence of higher
multi-pole moments over the dipole moment relative to the previous 1996 solar
minimum. The large difference between these two parameters is probably due to a
very low (three times lower than that in the 1996 solar minimum)
magnitude of the dipole component.

The observed diminution of the polar magnetic field also implies reduction of
the dipole component relative to the rest of the multi-pole components in the
solar magnetic field harmonic spectrum (Luhmann et al.\ 2009). This, in turn,
might result in re-enforcement of the multi-pole magnetic configuration in the
global solar magnetic field, which can be observed as persistent CHs along the
low- and mid-latitudes on the Sun.

From September 2006 until May 2009, there were at least five long lived CHs
associated with patches of open field inside the near-equatorial $\pm$40$^\circ$
zone. Three of them were of positive magnetic polarity and the other two were
negative polarity CHs. They lasted from 7 to 27 solar rotations. The longest
lived coronal hole, N2, formed on the solar disk during a time interval of
nearly 2 days and existed for 27 rotations at solar latitudes between S10 and
N08. N2 was associated with a fast solar wind stream at 1 AU of approximately
620$\pm 40$ km s$^{-1}$, which may in part be responsible for the heavy tail in
the solar wind speed distribution (Luhmann et al.\ 2009). An MHD model of the
corona that used SOHO MDI synoptic maps from June 25--July 21, 2008 for the
boundary condition, predicted an open field structure and dark emission in the
N2 CH region.

A possible mechanism for the formation of this global solar multi-pole is a new
challenge for solar dynamo theory. How do these persistent low- and mid-latitude
CHs form during a deep solar minimum? They are most likely not transient CHs
because there were no coronal mass ejecta during these quiet times that could
have created them. It is also very unlikely that these CHs separated from the
polar CHs and migrated to the solar equatorial region during one rotation
(see Section 3).

Very often a low latitude CH originally forms in association with an active
region (e.g., Bohlin \& Sheeley 1978;  Harvey \& Recely 2002; Bilenko 2004; 
Karachik et al.\ 2010). Recent analysis of CH formation (Karachik et
al.\ 2010) showed that appearance of a CH in association with an AR is not
unique. Authors discuss four cases observed during 2009 when an isolated CH
appeared on the place of decaying active regions (similar to the N2 CH case). 
These isolated CHs might form {\it in-situ} as a result of a large-scale
reconnection between the remnant fields of an decaying active region and
strongly diverged global open field lines (Wang \& Sheeley 2004).
Then as the active region decays the CH may or may not survive --
depending on the rate of AR field decay, whether there is new flux emerging in
the same place (e.g. a long-lived "nest" or "active longitude"), or just an
extremely slow decay of the AR fields, perhaps due to very little new flux
emergence going on. Anyway, the isolated CHs, being very persistent during the
current 2007 solar minimum, contribute significantly into the total solar
magnetic field balance, equilibrium between the multipole and dipole
magnetic components and, eventually, might play a role in the onset of the next
solar activity cycle. 

We are thankful to A. Pevtsov and G. deToma for very helpful discussions. 
We highly appreciate the efforts of the anonymous referee whose comments and
criticism led to improvement of the manuscript.
V.
Abramenko and V. Yurchyshyn were supported by NSF ATM-0716512 and NASA STEREO
NNX08AJ20G grants. J. Linker and Z. Miki\'c were supported by the LWS Strategic
Capabilities program (jointly funded by NASA, AFOSR, \& NSF), NASAs Heliophysics
Theory Program, the NSF CISM program, and the STEREO IMPACT and SECCHI teams.
Computational resources were provided by the TACC on the Ranger supercomputer
and by NAS on the Columbia and Pleiades supercomputers.
We thank the ACE MAG and SWEPAM instrument teams and the ACE Science Center 
for providing the ACE data. 
We acknowledge the Wilcox Solar Observatory staff for providing the harmonic
coefficients.
SOHO is a project of international cooperation 
between ESA and NASA. The SECCHI data used here were produced by an 
international consortium of the Naval Research Laboratory (USA), Lockheed 
Martin Solar and Astrophysics Lab (USA), NASA Goddard Space Flight Center 
(USA), Rutherford Appleton Laboratory (UK), University of Birmingham (UK), 
Max-Planck-Institut for Solar System Research (Germany), Centre Spatiale de 
Liege (Belgium), Institut d'Optique Theorique et Appliquee (France), and 
Institut d'Astrophysique Spatiale (France).

{}

\end{document}